\documentstyle[aps,prl,twocolumn]{revtex}

\input{psfig.tex}

\begin{document}

\newcommand\beq{\begin{equation}}
\newcommand\eeq{\end{equation}}
\newcommand\bea{\begin{eqnarray}}
\newcommand\eea{\end{eqnarray}}

\def\rb{\overline{\rho}}
\def\ro{\delta\rho}
\def\zb{\overline{Z}}
\def\zo{\delta Z}
\def\zgb{\overline{\ln Z_G}}
\def\zgo{\delta \ln Z_G}

\def\ab{\alpha,\beta}

\def\p{\rm p}
\def\s{\rm s}
\def\v{\rm v}
\def\w{\rm w}

\def\Avw{A_{\v\w}}
\def\Lvw{L_{\v\w}}
\def\fvw{f_{\v\w}}
\def\phivw{\phi_{\v\w}}
\def\thetav{\theta_{\v}}

\topmargin0cm
\draft
\twocolumn[\hsize\textwidth\columnwidth\hsize\csname@twocolumnfalse%
\endcsname
\title{The Weyl Series and the Trace formula: Can we add them?}
\author{R. K. Bhaduri, N. D. Whelan}
\address{Department of Physics and Astronomy, McMaster University,
         Hamilton, Ontario L8S 4M1, Canada}
\author{M. Brack$^1$, H. G. Miller$^2$, M. V. N. Murthy$^3$}
\address{$^1$Institute of Theoretical Physics,
             University of Regensburg, D-93040 Regensburg, Germany\\
         $^2$Department of Physics, University
             of Pretoria, Pretoria 0002, South Africa\\
         $^2$Institute of Mathematical
             Sciences, Chennai 600113, India}

\date{\today}

\maketitle
\begin{abstract}
Periodic orbit expressions for the density of states lead to spurious
results when directly used to calculate quantities of thermodynamic
interest. This is because the trace formula is usually valid only for
large energies while the calculations make use of it at all energies.
We present a prescription for circumventing this problem by isolating
contributions which arise from the inaccurate low-energy behavior, and
are spurious, from other subdominant contributions which are physical
and are not contained in the Weyl series. The method is tested by
analyzing the fermionic grand partition function for a disk billiard
and may readily be extended to other dynamical systems.
\end{abstract}

]

In many-body systems such as atomic nuclei, metal clusters, quantum
dots and traps with a dilute gas of bosons, much of the physics may be
unraveled from the quantum description of the single particle motion
in an appropriate mean field. Much effort has gone into developing
intuitive, semiclassical descriptions of single-particle quantum
dynamics \cite{Gutz,Balian0,Balian1,Berry0,Brack}. One ubiquitous
concept is the division of the density of states into two components,
each with a distinct semiclassical interpretation. The first varies
smoothly with energy and other parameters and is related to the
geometrical properties of the classical phase space; for billiard
systems this is the well-known Weyl series. The second component is an
oscillatory function and is related to the dynamics of the classical
system through various trace formulas. Balian and Bloch \cite{Balian1}
showed that the decomposition can be made exact at the price of
expressing the result in terms of certain integral representations. In
practice, this is not very helpful and one is led to the expedient of
expanding these out as asymptotic series. This leads to the Weyl term
plus an infinite sum over periodic orbits, each one of which has its
own asymptotic expansion in powers of $\hbar$. (Explicit higher order
terms in the trace formulas for more than one dimension have been
worked out only in a few trial systems \cite{Brack,Gaspard}.)  We will
refer to the sum over the periodic orbits to leading order in $\hbar$
as the trace formula. In the resulting approximation, however, it is
not immediately clear to what extent this leads to spurious results in
integrated quantities derived from the trace formula.  The focus of
this paper is to investigate this question, given a representation of
the density of states as an infinite asymptotic Weyl series plus 
the trace formula.  Certainly, this is the form in which the
semiclassical results are most commonly expressed and used
\cite{Gutz,Brack}. We would like to know to what extent we can
meaningfully use this as it stands: for example in using it to
calculate quantities of thermodynamic interest by integrating over it
with various weight functions. The short answer is that we can
not. The Weyl formula will typically have terms which cannot be
integrated directly from zero energy. This problem is, however, easily
circumvented by judicious integrations by parts \cite{long}. A more
serious problem, investigated here, is related to the fact that the
trace formula is usually inaccurate at small energies and can lead to
grossly incorrect results for the integrated quantities. We will show
that this problem can be overcome by identifying a spurious component,
which is due to the incorrect low-energy behavior. There is another
component that is subdominant but contains the real, physical
information about the periodic orbits.

In this paper we focus on the disk billiard, in part because its Weyl
expansion is known to very high order \cite{Berry2} and also because
of its physical importance, for example in quantum dots
\cite{qdots}. However, at the end we mention analogous results for a
variety of systems in both two and three dimensions. For the disk, it
is shown in \cite{Berry2} that the divergence of the higher-order
terms in the Weyl series is controlled asymptotically by the shortest
accessible periodic orbit through the phenomenon of resurgence
\cite{Dingle,Ecalle}. The intricate link between the trace formula and
the Weyl series has also been studied by Cartier and Voros
\cite{Cartier,Voros}. Another interesting system to study in this
regard is the harmonic oscillator \cite{long} where there is an exact
trace formula. Since it can be found by resummation of the Weyl
series, we can work with either form. There are no spurious terms
although there are still contributions from the periodic orbits which
are exponentially small in temperature, as we will find for the disk.

We begin with the following decomposition of the quantum density of
states: $\rho(E)=\sum_n\delta(E-E_n) \approx \rb(E) + \ro(E)$. The
Weyl part $\rb(E)$ is given in terms of reciprocal powers of $E$ as
well as delta functions and their derivatives. (The corresponding
expression for the partition function is simpler and is presented
below.) The trace formula for the oscillating part $\ro(E)$ (with
$\hbar^2/2m=1$ and disk radius $R=1$) is given by \cite{Brack,Reimann}
\beq
\ro(E) \approx \sum_{\v=2}^\infty\sum_{\w=1}^{[\v/2]}
{\Avw\over {E^{1/4}}}\sin(\sqrt{E}\Lvw+\thetav),
\label{rhoz}
\eeq
valid for $E\gg 1$. Each pair of integers $(\v,\w)$ represents a
one-parameter family of orbits, related by continuous rotations, where
$\v$ is the number of vertices (i.e., reflections from the boundary of
the disk) and $\w$ is the winding number. $\Lvw=2\v\sin(\pi\w/\v)$ is
the length of an orbit, $\Avw=\fvw\Lvw^{3/2}/\sqrt{8\pi}\v^2$ its
amplitude and $\thetav=(3\pi/4-3\v\pi/2)$ its phase. The degeneracy
factor $\fvw$ is unity for $\v=2\w$, otherwise it equals $2$.  We now
consider using this expression to calculate physical observables of
interest. As an example we consider the fermionic grand partition
function \cite{BM} for non-interacting particles, from which all
thermodynamic quantities can be derived:
\beq
\ln\,Z_G(\ab)=\int_0^\infty  \rho(E) \ln
\left[1+\exp (\alpha - \beta E)\right]\, dE\,.
\label{zg}
\eeq
Here $\beta$ is the reciprocal temperature and $\alpha/\beta=\mu$ is
the chemical potential. However the considerations of this paper will
apply to any quantity expressible as an integral over the density of
states, such as the total energy of a many-fermion system or the
integrated density of states.

To proceed it is convenient to introduce the one-body partition
function
\beq
Z(s)=\int_0^{\infty} \rho(E) \exp(-sE)~dE\,.
\label{aone}
\eeq
The parameter $s$ can be thought of as an inverse temperature for the
one-body problem. In what follows we will allow it to be complex, and
for the present purpose it is just used as a mathematical variable.
Noting that the {\it two-sided} Laplace transform of
$\ln\left[1+\exp(\alpha)\right]$ is $\pi/s\sin(\pi s)$ and
using the convolution theorem \cite{van} we can rewrite Eq.~(\ref{zg})
in the form
\beq
\ln Z_G (\ab)
= \frac{1}{2\pi i}
\int_C ds \,e^{\alpha s} \left\{ {\pi\over s\,\sin(\pi
s)} Z(\beta s)\right\}.
\label{hawa2}
\eeq
The contour $C$ runs from $-i\infty$ to $i\infty$ with the real part of
$s$ between 0 and 1, although obviously this can be deformed within the
rules of complex calculus. The study of the grand partition function
has the nice feature of being directly related to physical observables.
It has also the following mathematical advantage. It is well known that
the trace formula for the single-particle density of states is
generally divergent. However the finite temperature mitigates this. The
number of orbits in the disk increases as a power law with orbit length
while the temperature causes an exponential suppression with
temperature. The net effect is a convergent sum.

Substituting $\rb(E)$ into (\ref{aone}), one obtains
\beq
\zb(s)={c_0\over {s}} +
\sum_{r=1}^{\infty}~{c_r\over{\Gamma(r/2)}}~s^{r/2-1}
\label{hawa6}
\eeq
with $c_0=1/4,~c_1=-\pi/4,~c_2=1/6,~c_3=\pi/256$, etc. Berry and Howls
\cite{Berry2} tabulate the coefficients up to $c_{31}$. This expression
for $\zb$ may be substituted in Eq.(\ref{hawa2}) to obtain the Weyl
series for the grand potential. We find
\bea
& & \zgb(\ab) = {c_0\over \beta}
\int_0^{\infty}dE \ln(1+e^{\alpha-E}) \label{hawali} \\
& & \, \quad \qquad \qquad + \; \sum_{n=1}^N {c_{2n}\over \Gamma(n)}
\beta^{n-1}{d^{n-1}\over d\alpha^{n-1}} \ln(1+e^{\alpha})
\nonumber \\
& & + \sum_{n=0}^N {c_{2n+1}\over \Gamma(n+1/2)} \beta^{n-1/2} {d^n\over
d\alpha^n}
\!\int_0^{\infty} \! dE {1\over {\sqrt{\pi E}}} \ln(1+e^{\alpha-E}).
\nonumber
\eea
This series is asymptotic and diverges if taken to infinite order; it
may therefore only be summed up to a maximum value $N$ of the index
$n$. This is shown in Fig.~1, where the difference
\beq \label{delta}
\Delta(\ab)=\ln Z_G(\ab) - \zgb(\ab)
\eeq
between the quantum and the Weyl value is plotted versus $N$ (circles)
for selected fixed values of $\alpha$ and $\beta$. For
$\mu=\alpha/\beta \gg 1$, $\Delta$ settles down to a plateau after a
few terms of the series, giving the optimum result. The asymptotic
nature of the series (\ref{hawali}) is apparent for $\alpha/\beta=1$
in the bottom of Fig.~1, where $|\Delta|$ is seen to increase rapidly
with the inclusion of higher order terms. (The same happens for the
other cases, but at higher values of $N$.)  We see from this figure
that the series yields a plateau at a nonzero value, indicating that
(\ref{hawali}) deviates significantly from the true quantum result.
This is reminiscent of the phenomenon discussed by Balian {\it et al.}
\cite{Balian2} where it is shown how the error in using a given
asymptotic expansion can be dominated not by the least term in the
expansion used but rather in neglecting some weaker subdominant saddle
contribution. Clearly what is missing here is the contribution from
the periodic orbits.

We proceed by substituting into (\ref{zg}) the oscillating density of
states as given by (\ref{rhoz}). This gives a spurious result.  For
example, in the limit $\beta=0$ we are left with simple Fresnel
integrals. Upon doing the sum over orbits we get the finite
contribution $(\ln 2 -\pi^2/24)\ln(1+e^\alpha)/4\sqrt{2}$. However, we
know that in this limit the Weyl contribution is exact and clearly
adding this amount would destroy the agreement. Therefore, we should
not take the prescription of substituting the oscillating density of
states into (\ref{zg}) too literally.

Rather, we proceed as before by first calculating the contribution to
the one-body partition function. Changing variables to $k=\sqrt{E}$ we
have the following integral to do for each orbit
\beq \label{integral}
\int_0^\infty dk\, \sqrt{k} \sin(kL+\theta) \exp(-sk^2)\,,
\eeq
where we have temporarily suppressed the $\v$ and $\w$ indices. We
decompose the sinusoid into two exponentials and then separately
analyze the two integrals. There is an end-point contribution from the
lower limit which can be determined from Watson's lemma \cite{Dingle}.
There is also a saddle at either of $k=\pm iL/2s$ which may or may not
contribute depending on the phase of $s$; there is a Stoke's phenomenon
when $\Im s = 0$. The final result valid for $|s|\ll 1$ is \cite{F11}
\bea
\zo_{\s}(s)
       & \approx & \, - \!\! \sum_{\stackrel{\v,\w}{\v\,\rm{even}}}
               \fvw{(-)^{\v/2}\over \sqrt{8}\v^2}
               \sum_{n=0}^{N_{\s}} {(4n+1)!!\over 2^{2n} n!}
               \left({s \over \Lvw^2}\right)^n \! , \label{dong1} \\
\zo_{\p}(s)
       & \approx & \sum_{\v,\w} {\fvw \over {4\v^2}} \Lvw^2
             {e^{-\Lvw^2/4s}\over s} e^{\pm i(\thetav-\pi/4)},
\label{dong2}
\eea
where the upper and lower signs in the phase of (\ref{dong2}) hold for
$\Im s \stackrel{\scriptstyle >}{\scriptstyle <} 0$. When $\Im s=0$,
one should take the mean of the two expressions \cite{Dingle}. The sum
over $n$ in (\ref{dong1}) is asymptotic and must be truncated at some 
maximum value $N_{\s}$ that depends on the orbit.
The series (\ref{dong1}) comes from the end-point analysis. When
substituted into (\ref{hawa2}), its first term is the unwanted
contribution at $\beta=0$ mentioned earlier. In fact, the entire series
is unwanted. It results in power series contributions to $\ln Z_G$, but
we have already shown in Eq.~(\ref{hawali}) that its power series is
correctly generated by the Weyl series alone.  Therefore this entire
series is spurious --- hence the subscript s --- and is neglected in
what follows. This will be justified more rigorously below.
The second contribution (\ref{dong2}), by contrast, comes from the
saddle-point analysis. We will see that it does contain the physical
information about the contributions of the periodic orbits --- hence
the subscript p. It is interesting to note that (\ref{dong2}) is
exponentially subdominant compared to (\ref{dong1}) and yet carries all
the relevant information. Higher-order saddle-point corrections to
(\ref{dong2}) have been worked out \cite{oddseries}, but there is no
reason to include them here because they correspond to the missing
higher-order corrections to the trace formula (\ref{rhoz}) which are
unknown.

The origin of the spurious series (\ref{dong1}) is the low-energy
region, where the trace formula is inaccurate.  Removing this
spurious series could be thought of as subtracting from (\ref{rhoz}) a
series of the form $\ro_{\s}(E)= \sum_{n=0}^{N_{\s}} a_n
\delta^{(n)}(E)$. Requiring that all moments of the corrected
trace formula be zero fixes the coefficients $a_n$ and leads to
subtraction of (\ref{dong1}) from $\zo(s)$. Another argument for
ignoring $\zo_{\s}(s)$ is as follows. The derivation of (\ref{rhoz})
assumes energy is large. We are therefore free to replace this
expression by any other which is asymptotically the same.  However,
the expansion (\ref{dong1}) is not invariant under such a change
(unlike (\ref{dong2})) and we therefore conclude that it cannot
contain any meaningful information.  It is even possible to find a
version of $\ro(E)$ whose spurious contribution is identically zero by
inverse Laplace transforming the physical expression
(\ref{dong2}). This gives Bessel functions \cite{long} whose
asymptotic expansions are precisely $\ro(E)$.  However for small
energy they behave more smoothly and, most importantly, do not lead to
spurious structure. We interpret the result as a regularised trace
formula.

We now calculate the contribution of the periodic orbits to the
fermionic grand partition function by substituting $\zo_{\p}(s)$ into
(\ref{hawa2}) (while omitting $\zo_{\s}$) and doing the contour
integral numerically. We call the result $\left(\zgo\right)_{\p}$ and
show it in Table~1 for a few cases together with $\Delta(\ab)$.
Clearly this analysis captures the part missing from the Weyl
series. We stress that this would certainly not be the case had we
included the spurious series. We also show this data in Figure~1 as
the dashed horizontal line. For comparison we also show the result of
substituting (\ref{rhoz}) directly into (\ref{zg}), doing the
integrals numerically and subtracting off the spurious series found
from substituting (\ref{dong1}) into the relation (\ref{hawa2}). The
results are shown as crosses where the horizontal axis represents
$N_{\s}$ in (\ref{dong1}). The asymptotic result is well captured by
$\left(\zgo\right)_{\p}$.

Thus, the contribution of periodic orbits to the grand potential
accounts for the difference between the quantum and the Weyl
calculations in a consistent manner, but only after discarding the
spurious contributions. It is interesting to note that the periodic
orbits represent a much more important contribution to $\ln Z_G$ than
to $Z$ itself, for reasons we do not have space to discuss here
\cite{long}. The integral (\ref{hawa2}) has saddles at $s=\pm i
L/2\sqrt{\alpha/\beta}$ and the resulting stationary phase analysis
leads to a known form \cite{Kolomiets} valid in the limit of large
$\alpha$. However, by doing the integral exactly we can relax this
constraint and therefore have a more general result. (Both approaches
fail, however, when $\mu$ becomes much smaller than one.)

This sort of analysis can immediately be generalized to other classes
of periodic orbits. For example, a typical billiard orbit contributes
to trace formulas as $AE^{n/4}\sin(\sqrt{E}L+\sigma\pi/2+n\pi/4)$.
Examples are diffractive orbits in two dimensions, isolated orbits,
the disk orbits considered here, diameter orbits in the sphere and
polygonal orbits in the sphere for which $n=(-3,-2,-1,0,1)$
respectively. In its contribution to $Z(s)$, we can identify a
component which comes from the end point and is spurious and another
which comes from a saddle point and is physical. The latter is
approximately
\beq \label{latter}
A\sqrt{\pi} \left({L\over 2}\right)^{n/2+1}
             {e^{-L^2/4s}\over s^{(n+3)/2}} e^{\pm i(\sigma+n)\pi/2},
\eeq
which agrees with (\ref{dong2}) for the case $n=-1$ with $\sigma\equiv
2-3\v$ and substituting in the amplitudes from (\ref{rhoz}). This can
also be extended to potential systems, scaling or otherwise, although
the former is somewhat simpler.

In summary, a careful asymptotic analysis of integrals over
approximate trace formulas reveals that the physically relevant
contributions from the periodic orbits are contained in exponentially
subdominant terms, whereas the dominant terms that come from the
inaccurate low-energy behavior of $\ro(E)$ are spurious and must be
discarded. Adding only the subdominant terms to the Weyl series gives
good numerical agreement with quantum-mechanical results.

This work was supported by NSERC (Canada), DFG (Germany), and by FRD
(RSA).  M. Brack and H. Miller acknowledge the warm hospitality of
McMaster University where much of this research was carried out. We
thank David Goodings, Avinash Khare and Donald Sprung for useful
discussions.

\begin{table}
\begin{tabular}{rrrrr}
$\alpha$ & $\beta$ & $\ln Z_G$ & $\Delta$\phantom{$\Delta$}
& $\left(\zgo\right)_{\p}$\\
\hline
10 & 1.0 &  4.2499 &  0.2880 &  0.2549\\
10 & 0.5 & 12.6489 &  0.2144 &  0.2100\\
10 & 0.1 & 97.0390 & -0.0075 & -0.0071\\
 1 & 1.0 &  0.0083 &  0.0148 &  0.0152
\end{tabular}
\caption
{\small The logarithm of the grand partition function, the
difference from the Weyl approximation and the contribution from the
periodic orbits (with the spurious part removed) for a selection of
values of $\alpha$ and $\beta$.}
\label{table1}
\end{table}

\begin{figure} [h]
\vspace*{-1.5cm}\hspace*{-1.1cm} 
\psfig{figure=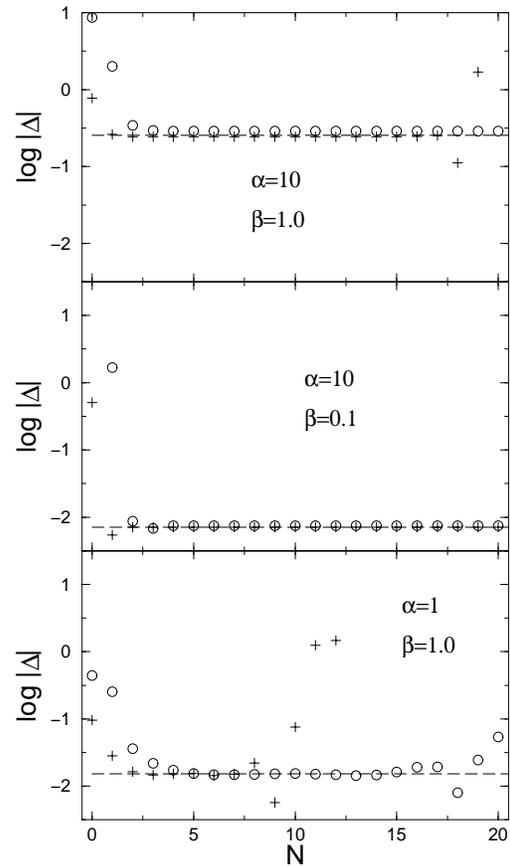,height=5.5in}
\vspace*{-1cm}
\caption
{\small The logarithm (base 10) of $\Delta$ versus the order in the
asymptotic expansion (circles). The dashed line is the periodic orbit
prediction $\left(\zgo\right)_{\p}$. The crosses result from
directly using the expression (\ref{rhoz}) in the energy integral
(\ref{zg}) and then explicitly subtracting off the spurious component
versus $N_{\s}$ in (\ref{dong1}).}
\label{figure1}
\end{figure}


\begin{references}
\bibitem{oddseries} Explicitly, Eq.\ (\ref{dong2}) is multiplied by the
asymptotic series $\left\{1-\sum_{n=1}^{N_{\p}}(-1)^n {(4n-3)!!\over {2^{2n}
n!}} \left({s \over L_{vw}^2}\right)^n\right\}$ (see \cite{long} for details).
\end{references}
\end{document}